\documentclass{sigchi}

\toappear{\scriptsize Permission to make digital or hard copies of all or part of this work for personal or classroom use is granted without fee provided that copies are not made or distributed for profit or commercial advantage and that copies bear this notice and the full citation on the first page. Copyrights for components of this work owned by others than the author(s) must be honored. Abstracting with credit is permitted. To copy otherwise, or republish, to post on servers or to redistribute to lists, requires prior specific permission and/or a fee. Request permissions from permissions@acm.org. \\
{\emph{DIS '20, July 6--10, 2020, Eindhoven, Netherlands.} } \\
\copyright~2020 Association for Computing Machinery. \\
ACM ISBN 978-1-4503-6974-9/20/07\ ...\$15.00. \\
http://dx.doi.org/10.1145/3357236.3395476}

\usepackage{balance}       
\usepackage{graphics}      
\usepackage[T1]{fontenc}   
\usepackage{txfonts}
\usepackage{mathptmx}
\usepackage{hyperref}
\usepackage{color}
\usepackage{booktabs}
\usepackage{textcomp}
\usepackage{graphics}      
\usepackage[T1]{fontenc}   
\usepackage{txfonts}
\usepackage{adjustbox}

\usepackage{mathptmx}
\usepackage{color}
\usepackage{booktabs}
\usepackage{textcomp}
\usepackage{comment}
\usepackage[normalem]{ulem}
\useunder{\uline}{\ul}{}
\usepackage{longtable}
\usepackage{wrapfig}
\usepackage{float}
\usepackage{csquotes}
\usepackage{enumitem}
\usepackage{multirow}
\usepackage{microtype}        
\usepackage{ccicons}          

\usepackage{todonotes}

\def\plaintitle{Visions, Values, and Videos: Revisiting Envisionings in Service of UbiComp Design for the Home}

\def\emptyauthor{}
\def\plainkeywords{Contravision; Design Fiction; Envisioning; Food Technology; Scenarios; Smart Home; UbiComp; Values in Design; Videos.}

\makeatletter
\def\url@leostyle{%
  \@ifundefined{selectfont}{
    \def\UrlFont{\sf}
  }{
    \def\UrlFont{\small\bf\ttfamily}
  }}
\makeatother
\def\pprw{8.5in}
\def\pprh{11in}

\definecolor{linkColor}{RGB}{6,125,233}
\hypersetup{%
  pdftitle={\plaintitle},
  pdfauthor={\emptyauthor},
  pdfkeywords={\plainkeywords},
  pdfdisplaydoctitle=true, 
  bookmarksnumbered,
  pdfstartview={FitH},
  colorlinks,
  citecolor=black,
  filecolor=black,
  linkcolor=black,
  urlcolor=linkColor,
  breaklinks=true,
  hypertexnames=false
}

\begin{document}

\title{\plaintitle}

\numberofauthors{1}
\author{%
  \alignauthor{Tommy Nilsson\textsuperscript{1}, Joel E. Fischer\textsuperscript{1}, Andy Crabtree\textsuperscript{1}, Murray Goulden\textsuperscript{2}, Jocelyn Spence\textsuperscript{1},\\ Enrico Costanza\textsuperscript{3}\\
    \begin{tabular}[t]{@{}c@{}}
  \affaddr{\textsuperscript{1}The Mixed Reality Laboratory}\\
  \affaddr{\textsuperscript{2}Horizon Digital Economy}\\
  \affaddr{University of Nottingham, UK}\\
  \email{\{firstname.lastname\}@nottingham.ac.uk}
  \end{tabular}\nobreak\qquad
  \begin{tabular}[t]{@{}c@{}}
  \affaddr{\textsuperscript{3}UCL Interaction Centre}\\
  \affaddr {University College London, UK}\\
  \email{e.costanza@ucl.ac.uk}
  \end{tabular}}
}

\maketitle
\begin{abstract}
  UbiComp has been envisioned to bring about a future dominated by calm computing technologies making our everyday lives ever more convenient. Yet the same vision has also attracted criticism for encouraging a solitary and passive lifestyle. The aim of this paper is to explore and elaborate these tensions further by examining the human values surrounding future domestic UbiComp solutions. Drawing on envisioning and contravisioning, we probe members of the public (N=28) through the presentation and focus group discussion of two contrasting animated video scenarios, where one is inspired by 'calm' and the other by 'engaging' visions of future UbiComp technology. By analysing the reasoning of our participants, we identify and elaborate a number of relevant values involved in balancing the two perspectives. In conclusion, we articulate practically applicable takeaways in the form of a set of key design questions and challenges. 
\end{abstract}




\section{Introduction}
Mark Weiser, by many regarded as the forefather of ubiquitous computing, famously envisioned a future with our everyday environments augmented through a wide spectrum of computational resources. Using the \textit{Sal} scenario to convey and illustrate his ideas, Weiser argued that such future UbiComp solutions ought to predominantly operate on our behalf calmly from the background, rather than requiring direct human control or oversight \cite{WEI99}. 

Much in line with Weiser's vision, recent years have seen a rise in popular on-demand 'just-in-time' services, aiming to increase the convenience of our everyday lives by providing transportation, accommodation, food delivery, and supporting mundane chores at the 'touch of a button'. Yet, a growing number of researchers have also been voicing their concerns that the development of such novel UbiComp solutions seems to be moving along a trajectory that is at odds with key human values, including the freedom to engage in tasks proactively and independently, to explore, discover, reflect and to socialize and collaborate with other people \cite{seely2006ubiquitous}. The tensions that exist between such human values, on the one hand, and the pursuit of ever more convenience and efficiency by virtue of new technology, on the other, are already well evidenced for example in the discourse around the effects of automation on the job market \cite{Fre17}. 

The domestic setting has been described as a leading growth area for emerging UbiComp solutions \cite{crabtree2003finding}, with the global value of the smart home sector set to exceed \$150 billion by 2024 \cite{markets}. In the light of such a trend, it seems inevitable that the impact of UbiComp on the everyday life will continue to increase, which in turn further amplifies the importance of rethinking our priorities when designing future solutions. As Yvonne Rogers explains, if the proliferation of ubiquitous computing is to be a success, we will thus need a \textit{"new set of ideas, challenges and goals to come to the fore and open up the field"} \cite[p. 418]{rogers2006moving}. 

In response, our paper presents an exploration of people's \textit{practical values} concerning emerging and future domestic UbiComp solutions; specifically, to anchor our enquiry in a relatable topic, we focus on technologies around food practices. In so doing, we aim to shed light on just what values people draw on when reasoning about these nascent socio-technical infrastructures. By drawing on envisioning \cite{REE12} and ContraVision \cite{MAN10}, we have developed two scenarios, each championing a different value-set: one set tilted towards convenience by virtue of calm computing, the other prioritising an active and social lifestyle.

In exposing our scenarios to 6 focus groups featuring a total number of 28 people, we present two contributions. The first one is substantive and elaborates the values that our participants draw on in their practical reasoning regarding how technologies perceptibly do or do not fit into their everyday lives. We present findings from a thematic analysis of the focus groups that organises people's reasoning around four emerging themes: convenience, trust, privacy and choice. Thus our paper contributes to a growing body of work on values in design (for an overview, see \cite{Fri13}). 



Our second contribution is methodological. We reflect on our approach that combines envisioning and ContraVision with a focus group approach in order to 'get at' people's own values. This in turn allows us to extend the ongoing discourse on the use of envisioning in design of future technologies.

\section{Related Work}
Our research is positioned in the context of domestic UbiComp technology (and critique thereof). Our analytic orientation, in turn, draws on work on values in design while taking on a distinct orientation towards \textit{practical values}. Below we offer a brief review of related literature. 

\subsection{Domestic UbiComp}
Drawing on the potential offered by technologies such as machine sensing, AI and wireless connectivity, a deluge of UbiComp solutions is currently emerging with the ambition to enhance nearly every facet of everyday life. A smart fridge capable of detecting when milk is running out and sending a notification to its user \cite{luo2009smart}, or a smart shoe measuring the number of steps taken and logging this information to the user's online fitness account \cite{kuniavsky2010smart} are just two examples illustrating this ongoing trend. 

With such solutions beginning to find their way into domestic environments, the  \textit{smart home} has become a familiar trope in the industry and academia \cite{harper2006inside}. Frances Aldrich defines the smart home as \textit{"a residence equipped with computing and information technology which anticipates and responds to the needs of the occupants, working to promote their comfort, convenience, security and entertainment through the management of technology within the home and connections to the world beyond"} \cite[p. 17]{aldrich2003smart}. 

A large number of domestic solutions and services that fit under this definition are either in the process of being rolled out or already in use. To cover such technological landscape holistically is understandably beyond the scope of this paper. Rather, in our own exploration of values concerning domestic ubiquitous computing, we anchor our interest in a relatable real-world topic in the form of proactive domestic food management technologies. 

Food has been noted for its pervasive nature in the domestic environment \cite{comber2012food}, making it into a topic that members of the public are generally able to relate to and reflect upon in user studies, such as those featuring envisioning and other forms of speculative design \cite{disalvo2012fcj}. Examples of technologies in this vein that have been elaborated by previous research include automatic drone deliveries \cite{o2017drone}, recipe recommending systems \cite{dolejvsova2018designing}, domestic food service robots \cite{de2011domestic} or diet personalization services \cite{dolejvsova2017soylent}.

Such emerging UbiComp solutions have been noted for their ability to improve the efficiency of existing services as well as enable new ones \cite{akyildiz2004wireless} by operating on behalf of the user. This is typically done through elements of agent-based and 'autonomous' computing in the sense that agents are employed as mediators of interaction between the user and the service provider, or other users. Research has suggested that such agent-based systems can help reduce work and cognitive load \cite{MAE94}, help to simplify interactions with complex technologies such as future energy infrastructures \cite{COS14,ROD13} and tariffs \cite{ALP17,FIS13}. Most closely related to our research, a recent study of a 'veg box' scheme has revealed the situated ways in which people integrate agency delegation into existing food practices \cite{VER18}. 

While UbiComp solutions may have the potential to do a lot of good in a domestic context, scholars have also been warning that the ongoing migration of smart technologies from workplace environments into the home may introduce a range of novel challenges. The fact that much of autonomous technology was originally designed with offices and other workplaces in mind, rather than people's homes, has prompted scholars to question the extent to which user requirements in these two distinct contexts overlap \cite{hindus1999importance}. Bill Gaver, for instance, argues that \textit{"there is a danger that as technology moves from the office into our homes, it will bring along with it workplace values such as efficiency and productivity at the expense of other possibilities"} \cite[p.1]{gaver2001designing}. In other words, Gaver argues that while efficiency represents a key priority in a work setting, we ought to not automatically assume the same has to apply in a domestic context. This suggests that designing acceptable solutions for the home may require a re-specification of user requirements and a refocusing of existing design goals and priorities. 

One such prominent attempt to refocus the design goals of ubiquitous computing  was presented by Yvonne Rogers, who argued in favour of re-prioritizing from 'proactive technologies' towards 'proactive people' \cite{rogers2006moving}. Instead of augmenting the environment and calmly serving the needs of the users, and thus reducing the need for humans to think for themselves about what to do, Rogers argued that we should consider how UbiComp technologies could be designed to \textit{"augment the human intellect so that people can perform ever greater feats, extending their ability to learn, make decisions, reason, create, solve complex problems and generate innovative ideas"} \cite[p. 411]{rogers2006moving}. Similar ideas have been echoed by a range of scholars. John Sealy Brown, for instance, has put forth an updated vision of UbiComp as a tool for catalyzing creativity by transforming the user from a passive participant to an 'active cocreator'. \cite{seely2006ubiquitous}. 

In the light of such arguments, we reason that domestic UbiComp technologies concerned with food management represents a particularly value-laden domain, challenging people to make choices on what we put in our bodies, where we shop, how much we spend, and where the food we buy comes from. Consequently, the food technology domain touches on a rich tapestry of value-sensitive topics, such as health and well-being, pleasure and pastime, cost of living, culture, sustainability, business, production or logistics. Prior work has shown, for example, that people draw on a complex range of contingent practices when making decisions about food shopping \cite{Hyl18}. In the following subsection we elaborate the notion of values in design further. 

\subsection{Values in Design}

\textit{Values in design} constitutes an emerging umbrella term encompassing approaches to systematically identifying and accounting for values in technology design \cite{flanagan2005values}. Such identification and elaboration of values (e.g. privacy \cite{cockton2018valorize}) to which designers, users, other stakeholders, and the surrounding society are committed is often seen as critical for converging on design requirements for future systems \cite{van2013translating}. 

Prior work that has considered values in design frames this paper's conceptual orientation; namely, our concern with how people's values shape their views on certain technological envisionings (and what can be learnt from this for design). The aim here is twofold: to review the  discourse on value-sensitive design familiar to HCI audiences, and to introduce the distinct orientation we take to frame our understanding of values as \textit{practical}, in that we understand values as part and parcel of reflexive accounts of everyday practice. 

The discourse on values in design within HCI can be traced back to the influential view in the social sciences that there is no such thing as a value-free technology, a position whose implications Anderson outlines: \textit{"(...)  since the design process is fundamentally about intervention to create change, it must be steered by value-orientations. If the values of the users of the technology are not dominant, then those of others (often those in whose interests it is to exploit or control the users) will be"}  \cite[p. 18]{And97}.

Friedman has advocated for value-sensitive design (VSD) as a means to consider human values \textit{"as understood from an ethical standpoint"} in the design of technology \cite[p. 22]{FRI96}. Sellen et al. elaborated this position further, arguing that the turn towards consideration of values in design is a natural progression in our changing relationship with computers, explaining that \textit{"...if in the past HCI was in the business of understanding how people could become more efficient through the use of computers, the challenge confronting the field now is to deal with issues that are much more complex and subtle"} \cite[p. 60]{SEL09}. 

The VSD discourse thus grapples with the intricate question of what a 'universal' value system might look like and how to address inherent conflicts, such as moral values being incompatible with personal or economic goals \cite{FRI96}. At the same time, Borning and Muller \cite{BOR12} urge for care in VSD not to overclaim, e.g., by avoiding claims of universality and overgeneralisation. 

While our work does not engage directly with morality per se, we take care to describe the values we have uncovered without stating unsupportable claims. In particular, we follow Borning and Muller's recommendation to strengthen the 'voice of participants' in that we seek to foreground their practical reasoning \cite{BOR12}. In other words, we seek to explicate the ways in which our participants bring to bear their own value judgements about future technology envisionings, and care less about the extent to which these values are 'universal', or morally 'right' or 'wrong' (cf., \cite{BOR12}) \footnote{We clarify our stance here merely to say that it is not our goal to moralise participants' judgements, not to say that we are above moral judgements ourselves; on the contrary, it is important to note that, as Jayyusi said, \textit{"one cannot get out of the moral order in order to talk about the moral order"} \cite[p. 247]{JAY91}.}. 

Our goal then is to generate insights that designers can draw on by examining the ways in which people make value judgements accountable in their own practical reasoning about technology. Thus, we take on an understanding according to which 'values' are matters of practical and ongoing relevance for members in the conduct of their everyday lives, as\textit{ "moral reasoning is practically organised, and at the same time, practical reasoning is morally organised"} \cite[p. 241]{JAY91}. For brevity, we refer to this orientation as concerned with \textit{'practical values'}. Following this view, we unpack the ways in which people offer value judgements reflecting their everyday domestic practices. We do this by examining the participants' accounts in and through which they reason about how envisioned future technologies fit into, or rub up against this practical action and reasoning.

\section{Approach}

In order to provoke people to draw on values in their reasoning about future UbiComp systems, we developed an approach centred around the use of scenarios \cite{CAR95}. Here, we present our underlying rationale by drawing on envisioning (e.g., \cite{REE16}), and ContraVision \cite{MAN10}.

\subsection{Envisioning and Design Fiction}
Our approach to creating scenarios borrows from 'envisioning'. The focus on the future and on envisioning hypothetical interactions with technologies has been described by Stuart Reeves as a \textit{"characteristic future-oriented technique for design thinking"}, and represents a particularly prominent driver of ubiquitous computing research \cite{REE12}. Reeves unpacks the ways in which envisionings mix fiction, forecast and extrapolation in examples such as Weiser's aforementioned 'Sal scenario' \cite{WEI99}. Similar to other envisionings of future technologies in UbiComp and HCI \cite{REE12,ROD13}, we developed our scenarios by drawing on existing enabling technologies, and extrapolating from current technological capabilities. Hence, our scenarios depict a plausible 'near future', and are thus distinct from design fiction, which tends to employ science fiction in design thinking as a way of imagining the future \cite{STE09}. 

Nevertheless, our goal to provoke expectancies (in the sense of 'calling forth') by exposing people to future scenarios does share commonalities with some design fiction work, such as Blythe's work on 'imaginary abstracts' of fictional research papers, which seeks to link design fiction more explicitly to research \cite{BLY14b}, and Cheon and Su's suggestion to employ 'futuristic autobiographies' as a resource to explore user's stance on prospective technologies \cite{cheon2018futuristic}. 
Similarly, Nathan et al. have proposed 'fictitious value scenarios' as an extension of scenario-based design to incorporate critical and wide-ranging sets of implications \cite{NAT08}, while Wong and Mulligan \cite{WON16} examine the use of corporate concept videos as a tool to help surface values relevant to imagined futures, such as who has power and agency with these technologies. 

Thus, we are not the first to employ envisioning to explicitly engage with values in design. However, one key aspect of our method that does set it apart is our reliance on ContraVision.

\subsection{ContraVision}
ContraVision is a technique to evoke a range of user reactions by developing and presenting visions that are deliberately contradicting each other \cite{MAN10}. Mancini et al. argue that conveying two contrasting yet comparable representations of the same technology can help elicit a wider range of reactions from prospective users than what could normally be gained by conveying only one perspective \cite{MAN10}. The authors claim that this approach represents a particularly valuable method when seeking to explore users'  responses to technology that does not yet exist in any usable form, and when there is reason to believe that said technology is likely to raise subtle and elusive personal, cultural and social issues that can potentially jeopardise its adoption.

\subsection{Developing Scenarios of Future UbiComp Systems}
By using the ContraVision method as a premise, we thus created two video scenarios conveying contrasting values in design, aiming to provoke people to present their own values about future  technologies. These video scenarios then formed the focal point of our engagement with users. 

The use of scenarios has a range of distinct advantages that made them into our tool of choice. Beyond simply being time and cost efficient \cite{CAR00}, scenarios are also recognised for their ability to bridge the gap between abstraction and detail by vividly envisioning innovative concepts and allowing users to experience them, which in turn promotes innovative thinking and helps raise relevant questions \cite{MAC00}.

While scenarios can be conveyed in many different forms, including writing or graphic storyboards, we chose to base our enquiry around videos. As Young and Greenlee explain, the use of video scenarios makes it possible to illustrate a design vision of prospective systems more effectively than written documents or static sketches \cite{young1992participatory}. Moreover, Muller argues that videos can help more accurately simulate the use of not-yet-developed tools and technologies to explore new possibilities and that they thus help facilitate a \textit{"fuller understanding by focus group members, leading to a more informed discussion" }\cite{muller2007participatory}. Video-based scenarios have likewise been noted for their unique ability to contextualize and provide a rich socio-technical backdrop for prospective UbiComp technologies \cite{nilsson2019creating}.

\def\arraystretch{1.3}
\begin{table}[h]
\small
\begin{tabular}{ |p{3.9cm}|p{3.9cm}|  }
\hline
\multicolumn{1}{|c|}{\textbf{Engaging computing}} & \multicolumn{1}{|c|}{\textbf{Calm computing}} \\
\hline
\multicolumn{2}{|c|}{Scene 1: Food ordering} \\
\hline

Paula is returning home from her work using public transportation. A mobile app presents her with a list of recommended meals that she might have for dinner. The recommendations are based on the ingredients that are in season and locally available. & Paul is returning home from his work in a self-driving car. While still in the car, Paul is contacted by his personal robotic assistant who enquires about his choice of meal for tonight. Paul chooses an expensive fish imported from across the globe.  \\
\hline
\multicolumn{2}{|c|}{Scene 2: Waste management} \\
\hline

At home, Paula is notified by her smart home system that some of the groceries in her fridge are approaching their use-by date. She is encouraged to either consume them or donate them to a local food bank. She chooses the latter. & Paul does not have to care about food waste. Any food item that has gone out of date is automatically disposed of by his personal assistant.  \\
\hline

\multicolumn{2}{|c|}{Scene 3: Food delivery} \\

\hline
Paula is notified that her food order has been delivered to a central collection point. She cycles to the delivery point, leaves her leftover food in a food bank, grabs her new food ingredients and returns back home.  & The food ingredients that Paul ordered are delivered by a drone straight to his doorstep. From there, everything is taken care off automatically by his domestic robots.   \\
\hline

\multicolumn{2}{|c|}{Scene 4: Food monitoring} \\

\hline
Once back at home, Paula manually scans her newly acquired food items into her smart home system. This is to keep the system up to date on the available food and its use-by dates. & Every food item in Paul's fridge is being monitored by cameras and sensors. Whenever a particular food is running out, Paul is notified by an automated message and the food gets automatically restocked.  \\
\hline

\multicolumn{2}{|c|}{Scene 5: Cooking} \\

\hline
Once all the food has been scanned into the system, Paula proceeds by cooking and enjoying the dish she was looking forward to, safe in the knowledge that she is cutting down her food waste and sharing with the local neighbourhood. & Paul is relaxing on his sofa while watching some TV. Any important notifications are delivered by a virtual agent straight from his television. Meanwhile, his home robot cooks and serves him the dinner he ordered earlier that day.  \\
\hline
\end{tabular}
\caption{An outline of our contravision scenarios.} 
\label{table:scenarios}

\end{table}

\begin{figure}[h]
\includegraphics[width=0.47\textwidth]{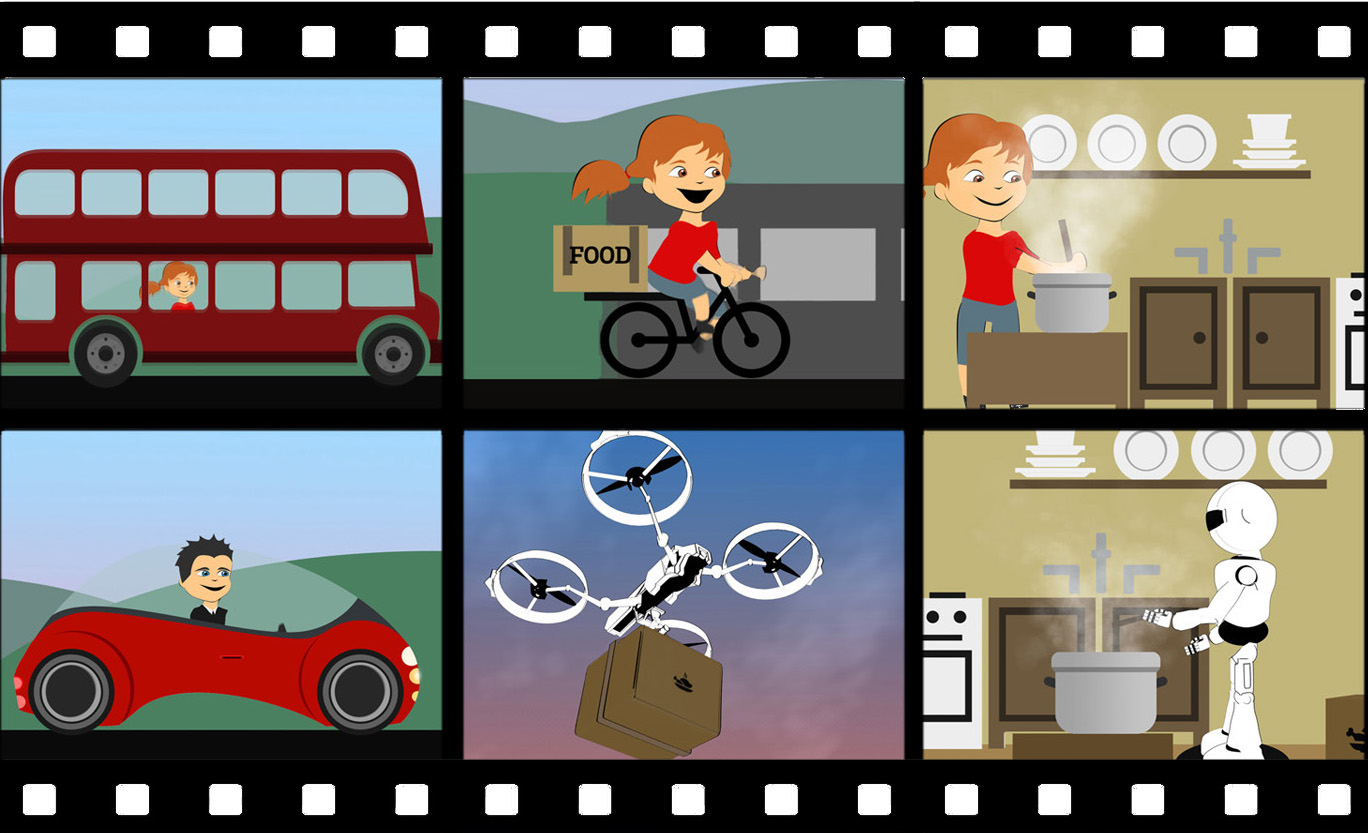}
\caption{A selection of scenes from our video scenarios.}
\label{fig:scenarios}
\end{figure}

More specifically, our video scenarios took on the form of cartoon-style animations (see Figure \ref{fig:scenarios}). This was largely for practical reasons, with cartoonish elements being easier and quicker to produce than more realistic visualisations. We reasoned there is nothing to be gained by faithful artistic renderings that in any case would not match any given participant's own home. 

The first scenario emphasised design values pertaining to the preservation of user's active lifestyle, with trade-offs in the form of less convenience. For instance, grocery orders were not delivered straight to the user's home, but rather to a central pickup point where they had to be collected. The scenario was grounded in familiar technologies such as mobile apps, grocery delivery services, 'click and collect' (online ordering and in-store collection popular in the UK), recipe recommendations, and barcode scanning (again, familiar to most people in the UK from 'self-checkouts' present in most supermarkets). Henceforth, this version of our scenario will be referred to as the \textit{engaging computing} scenario.\footnote{Full video scenario available at https://vimeo.com/238701035}

The second scenario encompasses values in design that prioritise the convenience of its users above everything else. In practice this meant that the envisioned user barely had to do anything, with food management and cooking being taken care of by systems operating in the background. The conveyed downside to such a solution was its high cost, not just in financial terms, but also in terms of having little interaction with the local community. Once again, we grounded the scenario in familiar technologies such as delivery services, supermarket loyalty programs, and recommendations, and included potentially more controversial ones such as drones and domestic robotics. Henceforth, this scenario will be referred to as the \textit{calm computing} scenario.\footnote{Full video scenario available at https://vimeo.com/241688455}

As evident in Table \ref{table:scenarios}, neither of the two scenarios is \textit{right}, as in \textit{correct}, and nor are they meant to be. Rather, our goal is to convey deliberately exaggerated visions of near future socio-technical ecosystems, and thus provoke our participants into reaction and reflection by 'breaching taken for granted background expectancies' that might otherwise be left unspoken \cite{nilsson2019breaching}. 


\section{The focus group study}
After obtaining ethics approval from the university's ethics committee, we hired an agency to recruit adults of all ages with diverse backgrounds (see Figure \ref{table:participants}). 28 members of the public eventually took part in our study (14 identified as female) as detailed in \autoref{table:participants}. Eleven participants were in their 20s, three in their 30s, seven in their 40s, three in their 50s, and four in their 60s. The median age was 39. Six had children aged 11 years of age or younger, three had a teen-aged child, and three had adult children. Our participants had a varied range of occupations; only four of them were students. All our participants were ethnic British. This is potentially significant due to the fact that values are frequently culturally dependent \cite{armenta2011relation}. We would therefore advise that any attempts at cross-cultural generalisation of our findings are approached with caution. 

\def\arraystretch{1.25}
\begin{table}[h]
\small

\begin{tabular}{ |c|c|c|c|c| } 
\hline
\textbf{Group} & \textbf{Participant} & \textbf{Gender} & \textbf{Age} & \textbf{Occupation} \\

\hline

\multirow{ 2}{*}{1} & P1 & Female & 25 & \hspace{0.9cm}Homemaker\hspace{0.9cm}  \\ 
& P2 & Female & 24 & Factory worker \\ 
& P3 & Female & 57 & Bar steward  \\ 
& P4 & Male & 47 & Careers adviser  \\ 
& P5 & Male & 64 & Medical advisor \\ 
\hline

\multirow { 2}{*}{2} & P6 & Male & 67 & Project manager \\
& P7 & Male & 47 & Tutor/community worker  \\ 
& P8 & Female & 19 & part-time student  \\ 
& P9 & Male & 24 & Personal trainer \\ 
\hline

\multirow{ 2}{*}{3} & P10 & Female & 55 & Private therapist  \\ 
& P11 & Male & 32 & Senior administrator  \\ 
& P12 & Female & 23 & Customer service  \\ 
& P13 & Male & 46 & Reactive engineer \\ 
\hline

\multirow{ 2}{*}{4}
& P14 & Female & 37 & Part-time student  \\ 
& P15 & Female & 26 & Student  \\ 
& P16 & Male & 49 & Trainer  \\ 
& P17 & Female & 36 & Manager  \\ 
& P18 & Male & 60 & Audit manager \\ 
\hline

\multirow{ 2}{*}{5}
& P19 & Female & 24 & Account manager \\ 
& P20 & Male & 49 & Student  \\ 
& P21 & Male & 65 & Civil servant  \\ 
& P22 & Male & 47 & Delivery driver  \\ 
& P23 & Female & 55 & Community food manager  \\ 
\hline

\multirow{ 2}{*}{6}
& P24 & Female & 25 & Recruitment  \\ 
& P25 & Male & 26 & CAD Technician  \\ 
& P26 & Female & 19 & Student  \\ 
& P27 & Male & 24 & Factory Packer \\ 
& P28 & Female & 41 & Solicitor \\ 
\hline
\end{tabular}

\caption{Focus group participants.}
\label{table:participants}

\end{table}

We adopted a semi-structured focus group approach to discuss the two contrasting scenarios with our participants. We chose this approach because group processes can help people explore and clarify their views in ways that would be less easily accessible in a one-to-one interview \cite{kitzinger1995qualitative}. Mays et al. argue that this is particularly the case when the researchers wish to encourage their participants to explore the issues of importance to them, in their own vocabulary, generating their own questions and pursuing their own priorities \cite{MAY96}. Similarly, Adams and Cox argue that focus groups\textit{ "can provide a flexible and participatory method that contextualises user's perceptions and experiences"} \cite[p. 33]{ADA08}. While focus groups run the risk of individuals taking over and dominating the discussion, we attempted to mitigate this through moderation so that each participant was able to express their opinion.

\subsection{Procedure}
The study was carried out over 6 two-hour long focus group sessions, each featuring 4-5 individuals. After providing informed consent on participating in the study, the focus group participants were initially introduced to the \textit{engaging computing} scenario, which was then followed up by the \textit{calm computing} scenario. After watching each scenario, participants were given room to reflect on what they saw. To cancel out any bias resulting from the order in which the two scenarios were seen, every other focus group had the order reversed. 

\subsection{Analysis}
All sessions were audio and video recorded. The group discussions were subsequently transcribed in full and served as a basis for a 'thematic analysis' \cite{BRA06}. The data was independently coded by two researchers, with inconsistencies subsequently addressed through discussion. The dataset was focused on the statements in which people made reference to their own lives; this served to index these as members' accounts pertaining to their everyday practices. Researchers then iterated over the coded dataset, grouping the data more broadly around the values that participants brought up in their practical reasoning offered in their statements. 

Note that the participant statements are often drawing on multiple values at the same time; values are not mutually exclusive. Note also the headlines we present are merely labels created through thematic analysis for presentational purposes rather than the participants' own categories. Thus, although we acknowledge that any grouping or labelling we perform will inevitably introduce some degree of our own interpretation, our thematic analysis strives to exhibit the participants' own value judgements (cf., \cite{CRA15}).

\section{Findings}
This section exhibits the ways in which our participants offered up practical reasoning about the envisioned scenarios. One of the striking aspects was how the majority of participants initially voiced a strong preference for the \textit{engaging computing} scenario, only to gravitate towards a significantly more nuanced view once relating the depicted ideas to their own everyday practices. Below we detail this process of participants drawing on their values pertaining to 4 key themes identified through our thematic analysis: convenience, trust, privacy and choice. 

\subsection{Convenience}
In the light of the contrasting levels of manual work depicted in the two visions presented to our participants, it came as no surprise that the value of convenience quickly established itself as one of the key focal points of the focus group discussions. While our participants generally agreed that convenience is often desirable, a number of predominantly social factors were likewise brought up suggesting that preserving some level of inconvenience would be equally important in the everyday domestic life. 

\textit{\textbf{P8: }I think again it's just not a very good thing to incorporate it [computing] into everything. I mean there's nothing… it's not that vital that you need to be told when something's gone off. It's not like you're gonna forget eating and cooking and you're gonna die. I just think like... if there's not a bit of inconvenience in our life... everything can't be like super easy, coz we need to be challenged.}

Our participants felt that technological solutions designed to calmly carry out our tasks on our behalf would in the same sweep also make our lives bland, monotonous and boring. There can, in other words, be such a thing as 'too much free time' or a 'too easy life'. 

Another concern brought up by our participants was that technology designed to enhance our convenience could as a side effect also lead to social isolation. An over-reliance on calm computing technology, they argued, might diminish the need for inter-human communication and in turn divest people of their social skills and erode the sense of empathy towards each other. As one participant stated, from a social point of view, the detriments of autonomous delivery services and other technologies could easily outweigh any of its benefits:

\textit{\textbf{P9:} If it turns up at your house and it's from a local source, potentially but... no... even for me, no. I like going out and chat, it's still what I gonna remember. It's like what... going back to the [calm computing] scenario, it's like... it's just making it bad interacting in real life. Social network does that already enough. It's like our kids are not able to talk to anyone, other than using their f***ing phones... it's just embarrassing.}

It is evident that for P9, the improvements to our lives brought about by novel technology have a downside in the form of negative effects on social interaction at large. While P9 along with a few other participants wholly rejected the idea of increased technology mediation, the majority acknowledged the importance of convenience and that addressing the concerns described throughout this section would render domestic calm computing technology acceptable. Similarly, while our participants generally looked positively upon solutions prioritising engaging system behaviour, they also agreed that such solutions would not be desirable unless also convenient to use. 

Overall, when reasoning to accept or reject proposed technological solutions, participants put forward the notion of convenience as a necessary condition. Requiring the user to do extra work, spend time, or go out of their way will make technology adoption unlikely. Convenience of use is necessary not for 'convenience's sake', but rather as it warrants technology use by increasing quality of life, such as to free up time for valued activities like spending more time with loved ones.

\subsection{Trust}

Given the concerns of social deprivation elaborated above, it might seem reasonable to conclude that, rather than automatically carrying out tasks on our behalf, future domestic UbiComp solutions ought to push users towards working collaboratively and thus foster deeper social ties in a community. This stance became particularly perspicuous when discussing the mechanisms of food sharing, as proposed by our \textit{engaging computing} scenario. Multiple participants suggested this might indeed be an opportunity to bring people together and strengthen a community. Sharing was in other words seen as a good way for building relationships and, in a sense, compensating for the loss of connection between neighbours that many people feel today. P10 shares her experience:

\textit{\textbf{P10:} We got, well, where we lived before, a couple of people on the street, they grew things in allotments. And there was often a random little bag of apples on the doorstep, you know. They just had an excess, just put it on your doorstep and shared it. That was really nice.}

However, as the discussions matured, it became clear that such sharing procedures represent a sensitive matter closely intertwined with the problem of trust. P26 elaborates:

\textit{\textbf {P26:} I think it's quite nice, like it's not nice wasting food if someone else wants it, like especially like homeless people, stuff like that. I've worked in like restaurants and stuff and they just throw away so much food, it's really sad. But obviously there's I think... with sharing your food, like I would feel maybe uncomfortable just getting random food from random people, coz I don't know if they've done something to it, like that kind of thing. But if I was giving my food away, I'd like that you're not wasting it.}

P26 here sums up the sentiment expressed by many of our participants; using technology as a means for engaging and coordinating homeowners to give away food, to avoid waste and thus benefit others in need is a welcome idea; however, being on the receiving end sparks discomfort. Many expressed that while they would be willing to give away their own unused food to others, they would not want to receive any themselves. 

\textit{\textbf{P28:} It would be better if [...], if you can give your food away that could go to charity or homeless shelter or something and you just pick up your own food. I think it's that bit, isn't it, for taking that seems to be the main problem.}

P28 here frames receiving food as 'the main problem', but suggests that  using a trusted entity such as a 'charity or homeless shelter' would go some way to alleviate the concerns. Some participants explicitly expressed the view that they 'felt uneasy' about the notion of being engaged and having to meet up with strangers to exchange food (P22). This sentiment was largely shared amongst our participants.

\textit{\textbf{P14:} I don't mind giving food out. I don't necessarily want food in. I'm not comfortable with that. I don't know whose house it's been in.}

\textit{\textbf{P17:} It is quite uncomfortable, isn't it? You don't know how it's been stored in the other house. I feel like cats and dogs... like some people allow their cats to walk inside and... I don't have animals at all...}

Most participants agreed that their willingness to accept a food sharing mechanism would largely depend on whether and how well they knew the person they were getting the food from. When dealing with strangers, mistrust was frequently displayed, as in the above exchange between P14 and P17, sparking a range of concerns regarding the manner in which the food had been stored and taken care of. 


One participant hypothesised a workaround solution in the form of a rating system, which would allow users to see the overall reputation of the person they were dealing with. This, he argued, could improve trust and in turn the overall user experience. 

\textit{\textbf{P25:} I think maybe if you took food off someone that... you know like when you go on Amazon or Ebay or whatever and someone's got 99\% rating. Like I'll take food off people that have a 98\% rating. Like if there was someone who had 60\%, would I take their cheese off them? I don't think so. You know what I mean, would I feel comfortable eating the cheese of a 60\% person? No. But 99\% person? Yeah!}

The attempt here by P25 to ameliorate trust issues more directly through technology, itself potentially highly problematic in its intrusiveness, emphasises the challenge of designing technologies for such sensitive exchanges. The proposed solution of rating opens up troubling questions about (anonymous?) public ratings of individuals. This is arguably more problematic even than the use of such trust proxies in commercial interactions (e.g., rating an Uber driver, where discomfort is mediated somewhat by judging someone in a work context, rather than them personally). Any domestic UbiComp solution designed to stimulate pro-social behaviour in this manner would consequently have to factor in values surrounding privacy.



\subsection{Privacy}
Another important topic raised in our focus groups while discussing calm and engaging system behaviour was the importance of privacy and accountability. Particularly, the discussion was occupied by the risks of privacy invasion as a consequence of having sensing technology integrated into home appliances (the scenarios both suggested the presence of monitoring technologies to track stock).

The prospect of being constantly monitored made the majority of our participants feel uneasy. One participant (P25) attempted to disperse these concerns by arguing that the monitoring technology proposed in our scenarios would in fact not mark a significant shift from the current state of affairs. As he explained, companies such as Google are already actively tracking our browsing history in order to present us with personalised recommendations prompting us to buy various products. P24 responded by arguing that there is a limit to how explicit and intrusive monitoring activities could be while remaining tolerable. Our calm computing scenario, for instance, depicted a virtually embodied assistant turning up on a user's TV in order to provide them with recommendations or reminders. P24 felt strongly that such a practice would represent a step too far: 

\textit{\textbf{P24:} Oh this is so scary. I wouldn't be comfortable with that to be fair. I mean the convenience is really, really good, like 'oh yeah you haven't had a fish for two weeks, maybe it's a good idea to have fish today' but then... like having all that data, like you know the Google (online shopping) is fine, coz that's just my shopping, but like the intake of my food, what's running out in my fridge, I think that's a bit personal. I think maybe if it was like um... giving someone a shopping list and then they deliver the shopping to your house, I think that would be fine, but turning up on your TV? [referring to the final scene] I know it's an exaggerated scenario but yeah... I think it's a bit too much.}

The domestic data collection and processing that enabled solutions such as recommendations in our scenarios were seen as intruding into matters that are 'a bit personal', participants felt uneasy being held accountable for actions they perform with food items (and other goods). Important here is that the level of intrusiveness seems specifically tied to certain objects, in that they accept that information from their shopping list may be used, but not if the same data is taken from their fridge. Also, in a similar vein, the view changes from delivery to the door (OK) to being on the TV (not OK). The reasoning here changes from what may be seen as the private space of the home (and the objects within), to the world outside and the objects acceptable for sharing with that world (such as a shopping list).

Going back to the notion of food sharing, our engaging computing scenario suggested that monitoring technologies could likewise be utilised to help coordinate exchange of food items between members of a local community. In spite of having good intentions, this idea also ended up generally rejected by our participants, who argued that sharing information about their food with others would make them feel self-conscious and judged by their peers.

\textit{\textbf{P17:} The fact that everybody... did it say something about everybody else knew your data in the community? Yeah, I didn't like that very much. Because I'd like the opportunity that if I can't be bothered to go and take that sour cream to the food bank and I want to put it in the bin then that's up to me.}

In other words, participants felt uncomfortable about someone else being able to see their potentially wasteful behaviour, specifically people 'in the community'. Technological solutions intended to minimise food waste could in this sense cause tension by putting pressure on people to justify their lifestyle. It is clear to see then that tensions emerge  here around issues of trust and community building. 

As the focus group progressed, it often became apparent that our participant's willingness to accept some level of monitoring in their homes was heavily dependent on the exact nature of the monitoring technology. A network of cameras was generally seen as invasive. Less revealing use of monitoring technologies was on the other hand received more positively. As P19 puts it: 

\textit{\textbf{P19:} I'd be happy with a camera getting specifics in my fridge or cupboard, but I wouldn't want any sound [capture] and I wouldn't want it put into a random room, coz you'd much rather have it in a room where it has to be specifically in a cupboard like a fridge, I would feel fine with it.}

It is evident that a sense of privacy is dependent on a multitude of locally organised factors, such as the physical implementation of the technology in this case. The nature of sensors, their location and the data being collected can make all the difference between acceptable and unacceptable. The recipient of the information intersects with these factors - whilst many would accept a service provider knowing the content of their fridge, the idea that people in their local area, particularly strangers, might also have access to this information was widely unpopular. Our participants, in other words, would require the ability to exercise free choice regarding the organization and routines in their home. 

\subsection{Choice}
While our calm computing scenario depicted technology providing its user with unrestricted choice of food items, the engaging computing scenario limited the choice to local and seasonal produce. Perhaps unsurprisingly, choice and autonomy became some of the most persistently discussed themes throughout our focus group sessions. Curiously, participants generally did not mind the limited choice that would come with a reliance on seasonal produce, explaining that the amount of choice available in current supermarkets is often excessive. 

\textit{\textbf{P10:} I find it overwhelming now, it's too much choice. I'd rather go to a small supermarket where there is less choice. It's time-consuming; you have to look at everything to make an informed choice.}

A frequently shared view was that reducing freedom of choice to make our lifestyles more sustainable or convenient would represent a welcome shift (P28). 


Although getting recommendations (e.g., based on the food that is in season or based on a user's diet) might in this sense be preferable, a number of participants also conceded that constraining choice too much would take away some flexibility (P14, P16). Furthermore, a technology making suggestions (e.g., by altering recipes in order to use up certain items that are 'going off' soon) could be seen as patronising as it implicitly tells a person what to do. One participant argued that technology should not 'take away the choice' to manage our own lives (P11). As the participant explained, he would rather decide for himself what to cook and when to discard food items. Similar arguments were brought up by other participants who argued that they are frequently cooking for themselves, and value doing so. 
Participants likewise emphasised the importance of being able to 'pick and choose' for themselves what fits with their own lifestyle and personal preferences. Too much recommendations and directions, they argued, would eventually deprive them of much of their freedom.  

\textit{\textbf{P3:} It's just takin'  over your brain, it's just telling you everything.}

A risk inherent in such systems then is that the system might not just be restricting freedom of choice, it might even be impacting people's autonomy by depriving people of independent reasoning. Cooking was brought up as one specific example of limited choice having a detrimental impact on the overall user experience.

\textit{\textbf{Interviewer:} But what about... because [the system] is also in your fridge, and it makes recommendations of stuff you could cook from that. So you would still do the cooking, but it would be giving you advice.}

\textit{\textbf{P24:} Then you' re not thinking anymore, are you?}

While our participants generally valued technology that aids user choice, particularly when it comes to making more sustainable or healthy choices, there is evidently a fine line between helping guide decision-making and constraining it with the effect of seeming to patronise the user.


Overall, our participants'  responses suggest that choice is fundamentally a question of agency and responsibility -- a desire for self-determination bound up with \textit{personal} freedom. Hence choice is welcomed as fostering agency, but 'too much' choice is rejected as inhibitive. Choice, then, can be seen as a burden, but at the same time, if taken away, this becomes threatening to selfhood, to one's role as an autonomous, proactive person.

\section{Discussion}
Herein, we turn to review and discuss our findings more broadly, revisiting some of the larger objectives of this work and engaging with the discourse in literature. 

\subsection{Understanding values in design through practical values}
In this work we have exposed participants to future domestic UbiComp scenarios intentionally imbued with conflicting values, trading off concepts that emphasise convenience and efficiency (in the \textit{calm computing} scenario) with those that prioritize a proactive lifestyle (in the \textit{engaging computing} scenario). Our study helped reveal the complex ways in which these values are encountered in and through the participants' practical reasoning. This in turn allows us to examine the tensions that exist between (the attempts at) value-sensitive design practice and peoples' actual lived experience. 

What our work offers to the broader discourse on values in design is thus a demonstration of a methodological  perspective we take on to frame our understanding of values as part and parcel of reflexive accounts of everyday practice, as Jayussi said \textit{"moral reasoning is practically organised, and at the same time, practical reasoning is morally organised"} \cite[p. 241]{JAY91}. This perspective commits us to explicating values from the point of view of 'ordinary members of society' \cite{Gar86}, a move that also may safeguard future researchers from potential problems pointed out in the VSD discourse, such as to overclaim or overgeneralise \cite{BOR12}. 

Consider, for instance, the tensions in our participants' orientation to the notion of food sharing. Underlying these tensions are, first, questions of trust, with individuals having close ties to the recipient being much preferred over strangers. Second, to use Goffman's terminology, is 'violation of territories of the self' \cite{GOF71}. Food, as something which we ingest into our bodies, is a rich source of symbolism around acceptable and unacceptable encroachments on our being. Often, Goffman points out, there is little ostensible logic in this, as in the case of people who share a drink from the same bottle but would never countenance eating the same sandwich \cite{GOF71}. The potential for food to carry such loaded meanings adds weight to the (absence of) trust in the source. Ultimately, despite participants'  interest in reducing waste and engaging with local community, the technologies proposed by our scenarios to do just this failed to successfully navigate their concerns. 

Instead of adopting the proposed calm computing alternative, in the form of efficient and automatic food management solutions, the result was one that our participants ended up advocating solutions that would enable them to donate unneeded food to a recognised - and therefore trusted - charity. This achieved the stated goal of avoiding waste and engaging with the local community by placing the exchange within a recognised social interactional framework. Thus, to achieve a trusted and trustworthy food sharing community, it appears that technology designers may face the need to identify and align with such recognised interactional frameworks. With them, they will inherit the need to deal with their complex value systems, and all of their tensions.

\subsection{Calmly engaging?} 

Let us now return to consider one of the stated goals of this work. We said that we wanted to provide insights that designers can draw on in the shaping of future technologies. It is worth recalling Cockton's passionate position on re-centring HCI around values, where he also reflected on the role of research: \textit{"Our role should be to understand what is valued by a system's stakeholders and support them in delivering this value"} \cite[p. 155]{Coc04}. And let's remember Anderson's point that design \textit{"...must be steered by value-orientations. If the values of the users of the technology are not dominant, then those of others (often those in whose interests it is to exploit or control the users) will be"} \cite[p. 3]{And97}. Towards this end, we sought to learn by unpacking the ways in which people make accountable in their practical reasoning the values in their orientations to future sociotechnical landscapes. But what have we learnt? 


In a sense our participant's reasoning shows that while people welcome values such as convenience and efficiency as drivers of technology design, this is preconditioned by the technology manifesting conformity insofar as people need to be able to integrate it into their practices of everyday life without causing too much disruption. This goes to the heart of the matter of trying to design calm systems for improving the user's convenience. To be viable, they should a) neither make more work for users, b) nor make them uncomfortable (e.g., by making them publicly accountable for food practices). 

Convenience and efficiency achieved through calm computing are thus in themselves not negative, nor should they be the focus of criticism. Rather, it is the cost at which they are achieved that ought to be the subject of scrutiny. Designing solutions that prioritise the convenience of their users can, for instance, help free up time for activities that are highly valued, such as enabling a user to spend more time with loved ones. In a different context, the same efforts might however give rise to negative implications, such as reducing a users' sense of self-determination and freedom of choice. 

Similarly, while our work confirms that people frequently value the notion of engaging computing and its potential role in community-building, more fundamentally the technology needs to fit in with people's practical needs as a prerequisite, such as the need for privacy. This echoes the important role of 'boundaries between personal and public space' \cite{kozubaev2019spaces} and Doryab et al.'s finding on peer-to-peer service transactions that \textit{"contextual convenience has a high impact on the acceptability rating of a service transaction recommendation"} \cite[p. 25]{DOR17}.

It appears then that values traditionally associated with calm computing, such as convenience and efficiency, do not necessarily constitute a polar opposite to those of a proactive and social lifestyle. By the same token, 'proactive technologies' and 'proactive people' do not by nature constitute mutually exclusive design goals. Thus, the visions of Weiser \cite{WEI99} and Rogers \cite{rogers2006moving} do not appear to be inherently irreconcilable after all. Rather, our findings suggest that the two can coexist, each with its own set of potential trade-offs emerging through their interplay with a rich palette of practical human values. 

Indeed, there is a notable connection to the promises of earlier homekeeping technologies: the washing machine was marketed with the promise to save time and labour, and yet the amount of work done in the home actually increased \cite{SCH83}. Edwards and Grinter reflect on the example of the washing machine in their critique of UbiComp technologies: \textit{"the washing machine encourages a critical perspective on whether smart home technologies are 'labour saving' or whether they (...) merely shift the burden of work"} \cite[p. 265]{EDW01}. Perhaps we should be critical again here and be cautious about the potential of ever more automation entering the home, and whether it actually 'saves time', or, as was the case with (many) earlier technologies, merely specialises, fragments, and shifts sites of labour elsewhere. Integrating new technology into the order of everyday routines itself requires appropriation work which can change 'the socio-temporal order of society' \cite{SHO03}.

What our findings show more fundamentally then is that there is a contextual interplay of values that mediate whether or not a technology is seen as potentially acceptable. These are invoked in relation to locally accountable practices, for example in rejecting the domestic robot as a companion, while embracing the potential benefit to eat more healthily as a result of a robot's labour. This is not to say that thinking about values in design is not worth doing. On the contrary, as Sellen et al. have pointed out, \textit{"the interaction between values and technology needs to be much more carefully navigated than before. (...) one set of design choices might highlight certain values at the expense of others. (...) the diversity, scope, and complexity of the technologies that HCI deals with make tradeoffs between values a conundrum, not a platitude"} \cite[p. 61]{SEL09}. We would suggest that one of the contributions of our work is that it demonstrates those tradeoffs in terms of the complex ways in which values are drawn on in reasoning about domestic UbiComp services, and that this can be instructive to sensitise designers. It is not enough for technology to support efficient or pro-social practices; if it is to be adopted, designers need to consider how practically convenient it is in use, whether it addresses other needs, and how it might make space for activities genuinely valued. 

At this point, then, rather than siding with either calm or engaging computing and articulating 'implications for design'\footnote{Dourish has argued the case very well for why our kind of research can be productive without explicit 'implications for design' \cite{Dou06}.}, we simply offer a range of questions that designers of future ubicomp technologies might find productive to consider:

\begin{itemize}[itemsep=-1pt]
    \item What values might people draw on to assess whether and how the technology could fit into their everyday lives? 
\item What issues might arise when thinking through how the technology could rub up against people's orientation to critical topics such as convenience, privacy, trust and choice? 

    \item What are the potential tensions between the values designers intend their designs to support and the values that people understand those technologies to support in the context of their daily lives? 

    \item Are there potential barriers to the adoption of the technology relating to value-based tensions? 
    \item What can you do as a designer to mitigate these tensions, and therefore address the potential barriers to adoption?

\end{itemize}

\subsection{Towards Making Envisioning Participatory}
Scholars have raised concerns about the balance of the representation of views and voices in the design of envisionings. Reeves, Goulden, and Dingwall have, for instance, called for more consideration for the ways in which visions influence design processes and research agendas \cite{REE16}. The authors argue that 'grand visions' and projections continue to shape subsequent technology development, for example in the UbiComp community. However, these visions may lack \textit{social legitimacy} in that they can be narrowly focussed, may not take into account the concerns of people outside of the UbiComp community itself or miss concerns from other parts of the world (cf., \cite{BEL07}). Specifically, Reeves \cite{REE12} has discussed the ways in which earlier envisionings, such as Weiser's Sal scenario, have oriented subsequent research, and therefore risk to \textit{"homogenise (...) technology research work towards a particular interpretation"} \cite[p. 1580]{REE12}. 

Our approach seeks to mitigate this concern through the creation of multiple scenarios based on different envisionings, rather than working under the implicit assumption of only one particular envisioning. We reason that by diversifying the conveyed vision, we open up greater space for the general public to mix, match and critique the envisioned ideas. Here we 'merely' sought to relay the general public's response to certain visions of future technology. Further participatory research is thus needed to address the lamented lack of 'social legitimacy' \cite{REE16} to involve the general public in shaping truly diverse and inclusive visions beyond our provocative scenarios.

Although members of the general public are not necessarily wiser about what the future might bring than designers are, as demonstrated by our research, they have their own value judgements of how technology might improve their lives. Thus, increased participation is likely to generate scenarios that are more socially legitimate than what experts alone can create. This is significant, for as Reeves et al. explain, a broader base of legitimised scenarios creates fertile ground for the design of better future systems \cite{REE16}.


\section{Conclusion}
In this work we have examined the practical values people draw on when reflecting on potential technologies surrounding their domestic life. In so doing, we have revisited the debate on visions as drivers of UbiComp technology research and development; in particular we examined how people respond to 'calm' and 'engaging' video scenarios imbued with contrasting and sometimes deliberately conflicting value-sets. Our findings suggest the pragmatic conclusion that the ideals envisioned by the two perspectives can coexist insofar as they are appropriated through peoples' everyday practices without causing too much disruption. Ultimately, it is not the implementation of any overarching grand vision then, that helps foster a passive or active lifestyle, but rather a contextual and locally organized interplay of practical values. We concluded by offering a set of reflective questions that may help designers locate and respond to issues around such values in their own design to create future technologies that may better fit into peoples' everyday lives. 

\section{Acknowledgments}
This work was supported by the Engineering and Physical Sciences Research Council (grant numbers EP/N014243/2, EP/M001636/1 and EP/M02315X/1). We are grateful to our participants and to Yvonne Rogers, whose keynote at the Halfway to the Future Symposium 2019 \cite{httf} has inspired a reframing of this work. 

\bibliographystyle{SIGCHI-Reference-Format}
\balance
\bibliography{references}

\end{document}